# Identification of organic molecules with a laboratory prototype based on the Laser Ablation-CosmOrbitrap

L. Selliez[1,2, (corresponding author)], C. Briois[1], N. Carrasco[2,8], L. Thirkell[1], R. Thissen[3], M. Ito[4], F.-R. Orthous-Daunay[5], G. Chalumeau[1], F. Colin[1], H. Cottin[6], C. Engrand[7], L. Flandinet[5], N. Fray[6], B. Gaubicher[1], N. Grand[6], J.-P. Lebreton[1], A. Makarov[9], S. Ruocco[2], C. Szopa[2,8], V. Vuitton[5], P. Zapf[6]

1-LPC2E, UMR CNRS 7328, Université d'Orléans, Cedex 2, France

2-LATMOS/IPSL, UVSQ Université Paris Saclay, UPMC Univ. Paris 06, Guyancourt, France

3-Laboratoire de Chimie Physique, CNRS, Univ. Paris Sud, Université Paris-Saclay, 91405, Orsay, France

4-Kochi Institute for Core Sample Research, Japan Agency for Marine-Earth Science and Technology, Kochi, Japan

5-Univ. Grenoble Alpes, CNRS, CNES, IPAG, 38000 Grenoble, France

6-LISA, UMR CNRS 7583, Université Paris Est Créteil et Université Paris Diderot, Institut Pierre Simon Laplace, France

7-CSNSM, CNRS/IN2P3, Univ. Paris Sud, Université Paris Saclay, Orsay France

8- Institut Universitaire de France (IUF), Paris, France

9- Thermo Fisher Scientific (Bremen) GmbH, Germany

Corresponding author:

Laura Selliez

Laboratoire de Physique et Chimie de l'Environnement et de l'Espace (LPC2E), UMR CNRS 7328,

3A, avenue de la Recherche Scientifique

45071 Orléans cedex 2, France

e-mail : laura.selliez@cnrs-orleans.fr

# Abstract

In the Solar System, extra-terrestrial organic molecules have been found on cometary primitive objects, on Titan and Enceladus icy moons and on Mars. Identification could be achieved for simple organic species by remote sensing based on spectroscopic methods. However *in situ* mass spectrometry is a key technology to determine the nature of more complex organic matter. A large panel of mass spectrometers has already been developed for space exploration combining different types of analysers and ion sources. Up to now the highest mass resolution reached with a space instrument is 9,000 at *m/z* 28 and corresponds to the DFMS-ROSINA instrument (Balsiger et al., 2007) dedicated to the study of the comet 67P/Churyumov-Gerasimenko's atmosphere and ionosphere, in a low pressure environment. A new concept of mass analyser offering ultra-high mass resolving power of more than 50,000 at *m/z* 56 (under high vacuum condition about $10^{-9}$ mbar) is currently being developed for space applications: the CosmOrbitrap (Briois et al., 2016), based on the Orbitrap™ technology.

This work challenges the use of LAb-CosmOrbitrap, a space instrument prototype combining Laser Ablation ionisation and the CosmOrbitrap mass analyser, to identify solid organic molecules of relevance to the future space exploration. For this purpose a blind test was jointly organised by the JAXA-HRMS team (Japan Aerospace Exploration Agency-High Resolution Mass Spectrometry) and the CosmOrbitrap consortium. The JAXA team provided two organic samples, whereas the CosmOrbitrap consortium analysed them without prior information. Thanks to the high analytical performances of the prototype and our HRMS data post-processing, we successfully identified the two molecules as HOBt, hydroxybenzotriazole ($C_6H_5N_3O$) and BBOT, 2,5-Bis(5-tert-butyl-benzoxazol-2-yl)thiophene ($C_{26}H_{26}N_2O_2S$), with a mass resolving power of, respectively, 123 540 and 69 219. The success of this blind test on complex organic molecules shows the strong potential of LAb-CosmOrbitrap for future space applications.

*Highlights:*

- Efficient ionisation of solid sample by nano-pulsed single laser shot
- Powerful analytical performances of CosmOrbitrap mass analyser
- Successful blind test identification of organics ionised by Laser-CosmOrbitrap
- Laser-CosmOrbitrap relevant technique for space exploration of organic rich worlds

# I) Introduction

The capability to study organic molecules and organic-rich environments in the Solar System is important for astrobiology (Horneck, 1995). This allows a better understanding of the chemical evolution that leads to the emergence of Life on Earth and provides constraints on the possible habitability of other planets or moons. Detection of organic molecules and traces of extinct or extant Life is therefore driving space exploration concepts for objects like Europa, Titan and Enceladus. Many space missions to these objects are either in preparation or proposed: the JUICE (JUpiter ICy moons Explorer) mission (Grasset et al., 2013), the Europa Clipper mission (Phillips and Pappalardo, 2014) or, among others, the Dragonfly mission (Principal Investigator E. Turtle, see the JHU/APL website for further details) selected in 2017 as finalist of the NASA New Frontiers program.

One of the best analytical tools used in space missions for chemical analysis is mass spectrometry. This technique allows to assess the elemental composition of environments studied. The detection and the identification of compounds lead to a better understanding of the chemistry occurring on diverse objects of the Solar System. A high diversity in terms of targets studied (planets, moons, small bodies etc.) and sample types (solid rocks, gaseous compounds in atmosphere, aerosols, ions etc.) is possible. In part, this is due to the multiple combinations of ion sources and mass analysers existing. Mass spectrometers have been boarded since the beginning of the space exploration and are still an instrument family essential to the future space missions. On the Cassini-Huygens mission, the instruments INMS (Ion and Neutral Mass Spectrometer) (Waite et al., 2004) or CAPS (CAssini Plasma Spectrometer) (Young et al., 2004) enabled the collection of a large amount of data and the improvement of our knowledge about Titan upper atmosphere (Waite et al., 2007) and Enceladus plumes (Waite et al., 2017). Thanks to modelling, attributions of a number of peaks in INMS data were successfully proposed, up to $m/z$ < 100. However, the mass resolution of these instruments (including INMS) did not enable to directly decipher the composition of the complex organic matter. That explains why the moons of Saturn are still highly requested in the discovery programs of space agencies. The best mass resolving power (m/Δm or MRP at Full-Width Half Maximum (FWHM)) of a space mass spectrometer, was provided by the DFMS (Double Focusing Magnetic Mass Spectrometer) instrument of the ROSINA (Rosetta Orbiter Spectrometer for Ion and Neutral Analysis) experiment on board the Rosetta mission with m/Δm 9,000 FWHM at *m/z* 28 (Balsiger et al., 2007). With these performances, the detection of species like $N_2$ and $O_2$ on a comet was made possible for the first time (Bieler et al., 2015; Rubin et al., 2015). In addition, a prebiotic compound, the amino acid glycine, at *m/z* 75, has been detected (Altwegg et al., 2016). Despite its high MRP, ROSINA/DFMS covered a mass range from 12 to 150 mass units (u) excluding the analysis of heavy organic molecules about hundreds of mass units.

It became a critical need to develop a new generation of mass spectrometers to go further in these studies, with high analytical performances required. In term of mass accuracy, a range of less than 1 to 5 ppm is needed in order to provide relevant molecular formula attributions. About the MRP, few tens of thousands (from 50,000 to 100,000) allow the separation of isobaric interferences at high *m/z* (up to *m/z* 500) of species with an exobiological interest. Detection and identification of complex organic molecules involved in chemical mechanisms occurring on Solar System bodies, such as the organic chemistry observed on Titan (Hörst, 2017) are thus possible. In the laboratory, the Orbitrap$^{TM}$ technology is now currently applied to the analysis of complex organic material of interest to space exploration, such as analogues of Titan's aerosols (Pernot et al., 2010) and soluble organic matter of meteorites (Bonnet et al., 2013; Gautier et al., 2016; Orthous-Daunay et al., 2013). It enables to cover a dynamic mass range, up to several thousands in mass units (Makarov, 2000), with a data acquisition time of about 1 second.

A mass analyser designed for space and based on the Orbitrap technology is currently being developed under the name of CosmOrbitrap (Briois et al., 2016), which would bring a technical breakthrough for direct *in situ* analysis. Mainly based on the analysis of metals with a LAb-CosmOrbitrap prototype, Briois et al., 2016 have shown the high analytical performances of this mass spectrometry technique, expecting this mass analyser to be part of a future space mass spectrometer. In this previous work, high performances have been achieved with a simple instrumental configuration of the prototype using direct laser ablation ionisation coupled with a CosmOrbitrap mass analyser through an Einzel lens. They obtained mass resolutions at FWHM of 474,000 on beryllium (*m/z* 9) and of 90,000 on lead (*m/z* 208), close to mass resolution of 60,000 at *m/z* 400 obtained with a commercial LTQ-Orbitrap XL instrument (Perry et al, 2008). Briois et al, 2016 have demonstrated a mass accuracy within less than 15 ppm over the 12 to 115 *m/z* range with the LAb-CosmOrbitrap prototype, rather larger than the 1-5 ppm given by commercial Orbitrap$^{TM}$ based instruments.

The present work focuses on the capability of the LAb-CosmOrbitrap to provide molecular identification, another important question in the development of a space instrument, by performing a blind test on two different molecules. Under a Japanese/French collaboration framework, two organic samples have been chosen by the JAXA HRMS team and sent to the CosmOrbitrap team without any information on their chemical composition. This work reports the blind analysis performed on these two "unknown" samples by the CosmOrbitrap consortium with the LAb-CosmOrbitrap prototype in order to identify them. Updates on the analytical performances of the LAb-CosmOrbitrap, on organics, are also given.

## II) Methods

### a) The laser ablation – CosmOrbitrap prototype

The blind test experiments have been conducted using a slightly modified version of the laboratory prototype previously described in Briois et al., 2016. To provide a better absorption of organic molecules in the UV resulting into a more efficient ionisation, the UV nitrogen laser at 337 nm was exchanged for a Nd-YAG laser at 266 nm (Goesmann et al., 2017). The latter is the "Brilliant" model provided by Quantel, with 4 ns pulse duration and about 100 µJ energy per pulse. The angle of incidence on the surface of the sample-holder is 50°, resulting in an elliptical shape of the footprint of the laser beam with a minor axis of typically 30 µm and a major axis of 40 µm (measured on silicon wafer) and energy density of 15 J.cm$^{-2}$.

The Orbitrap mass analyser cell commercialised by the Thermo Fisher Scientific (Bremen, Germany) is an ion trap in which ions are oscillating in a confining quadro-logarithmic electric field produced by barrel-shaped electrodes. The geometry of the cell and the injection of ions by electrodynamic squeezing was developed by A. Makarov (Makarov, 2000). The Orbitrap cell consists of two external electrodes, one central electrode and a deflector electrode. External electrodes are kept at the ground potential. Ions are injected in the cell after 1.1 kV acceleration, while a transition of the high voltage (from nominal voltages -2500 V to -3500 V) is applied to the central electrode and (from 0 to 350 V) to the deflector electrode. All experiments have been carried out in positive ion mode. We note that negative ions can also be studied by reverting the high voltages on the source, the central and the deflector electrodes. Due to the quadratic potential in the longitudinal direction, trapped ions oscillate at a frequency proportional to their mass to charge (*m/z*) ratio. The pulsation ω (in rad/s) and the *m/z* ratio are linked by the formula: $\omega = (k*(z/m))^{1/2}$, where k is a parameter depending on the shape and the voltage applied on electrodes (Makarov et al., 2009; Perry et al., 2008). This oscillation creates a perturbation of the potential difference between the external electrodes, providing the signal analysed by the instrument electronics.

Figure 1 represents the laboratory test bench used in this study. The Nd-YAG laser at 266 nm is used for ablation and ionisation. The prototype is composed of two vacuum chambers. The first one contains the sample at $10^{-8}$ mbar. Solid samples can be analysed, as they are usually pressed or dropped on the metallic surface of a small sample-holder (8 mm height, 7 mm diameter) itself screwed at the extremity of a rod. The second vacuum chamber contains the Orbitrap cell and is maintained at a pressure of $10^{-9}$ mbar. Both vacuum chamber are linked by a differential pumping system. The aperture between both vacuum chambers is smaller than 2 mm. Two antennas, one on each external electrodes of the Orbitrap cell, are directed to the pre-amplifier. The whole configuration is named LAb-CosmOrbitrap, standing for Laser Ablation CosmOrbitrap. As defined in Briois et al., 2016, the CosmOrbitrap **space** mass analyser includes the following elements (in red in Figure 1): the Orbitrap cell itself and its retaining ring, the ultra-stable high voltage power supply, the pre-amplifier and the data acquisition / command control board cards. These elements, composing the CosmOrbitrap, are involved in a TRL (Technology Readiness Level) development. TRL is a scale used by space agencies, such as NASA, with 9 levels. The first one (TRL 1) is the basic principle, an idea of a concept and the last one (TRL 9) a real system, on-board and operational. CosmOrbitrap elements are at an intermediate level, TRL 3, indicating they are laboratory elements set up as a proof-of-concept.

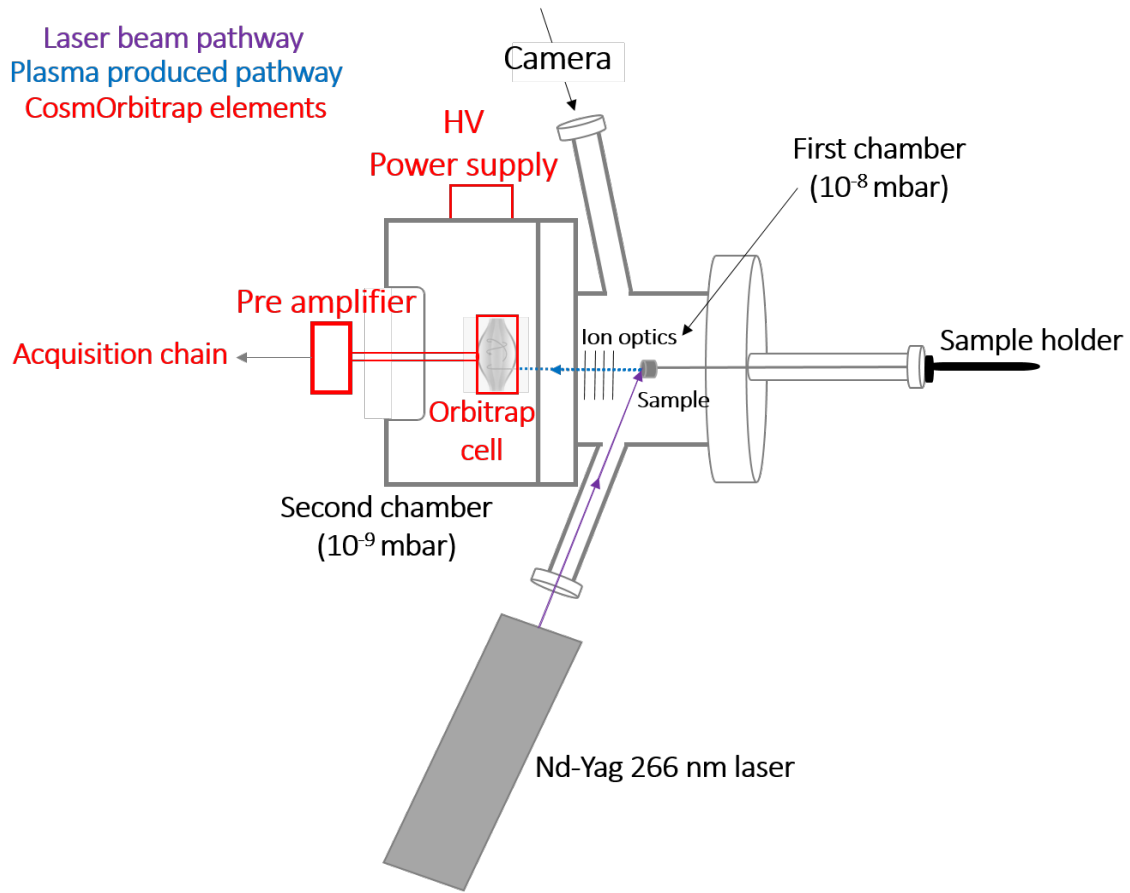

*Figure 1: LAb-CosmOrbitrap laboratory prototype at LPC2E (Orléans). The Nd-YAG laser is used as ablation/ionisation process. The first vacuum chamber holds the sample for ablation and ions extraction. The second chamber contains the Orbitrap cell for the mass analysis. Both chambers are connected by a 2 mm feedthrough. In red are reported the CosmOrbitrap elements involved in a TRL development for space applications. In the present work, CosmOrbitrap elements are at TRL 3.*

The acquisition software (Alyxan) allows visualising in real time both the frequency transient signal and the mass spectrum transient signal processed by Fast Fourier Transform (FFT). The signal is recorded during 838 ms with a sampling frequency of 5 MHz inducing the storage of 4 192 304 points, for a single packet of ions. A Hann window is applied before the FFT treatment.

### b) Sample preparation

Few milligrams of two samples, named A and B, have been supplied by JAXA HRMS team to CosmOrbitrap team. The samples are respectively white and green sticks (see Figure 2 for sample A).

A few sticks of each sample are collected and put on a different indium sample-holder. Sticks are pressed using an agate pestle to be embedded onto the indium target (Goodfellow, high purity: 99,999%, rolled) previously cleaned under consecutive ultrasonic baths of acetone and n-Hexane. The indium targets and the sample deposit are fed into the first vacuum chamber (Figure 1).

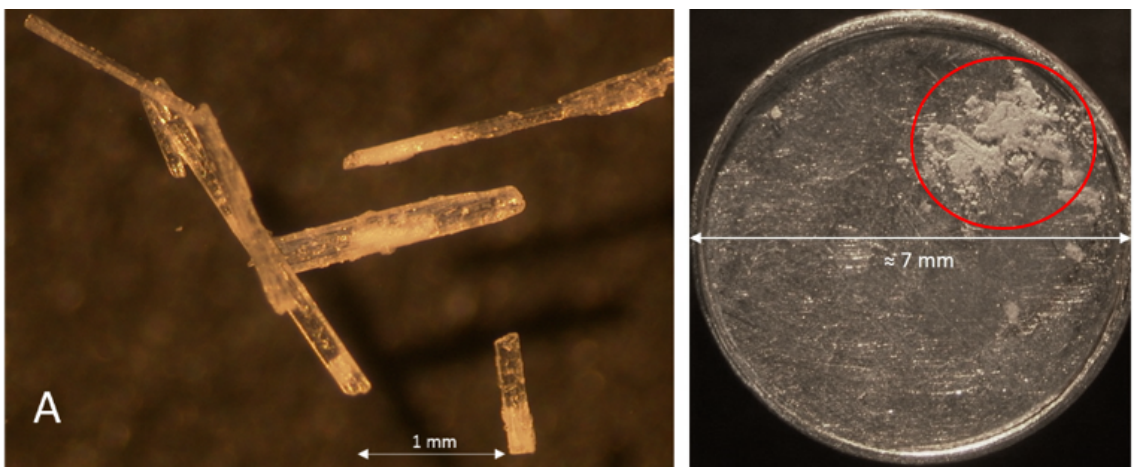

*Figure 2: Pictures obtained with an optical microscope of Sample "A" as it was received from the JAXA HRMS team (left) and pressed on the sample holder (right, in the red circle).*

This simple sample preparation provides a useful internal reference for mass calibration with the positive ion of indium major isotope peak at *m/z* 114.903, and possible cluster ions between the sample (M) and the indium ion as [In+M]$^+$, [In+2M]$^+$, [2In+M]$^+$, which help for the identification of the parent mass M. These clusters are also observed with other mass spectrometry techniques (Bhardwaj and Hanley, 2014; Carrasco et al., 2016; Le Roy et al., 2015) and with the LAb-CosmOrbitrap on another organic molecule, adenine ($C_5H_5N_5$) observed at *m/z* 135.0545 (unpublished data from the study of Briois et al., 2016). Adenine is studied in pure solid form (thin white powder) and Figure 3 shows a detail of a mass spectrum obtained from the analysis of adenine powder pressed onto an indium surface (as it was done for samples A and B studied in this work). Both indium isotopes ($^{113}$In$^+$ and $^{115}$In$^+$) are visible, as well as the adenine protonated peak [M+H]$^+$. Cluster of the molecular ion of adenine associated with indium [In+M]$^+$ is observed at the nominal *m/z* 250.

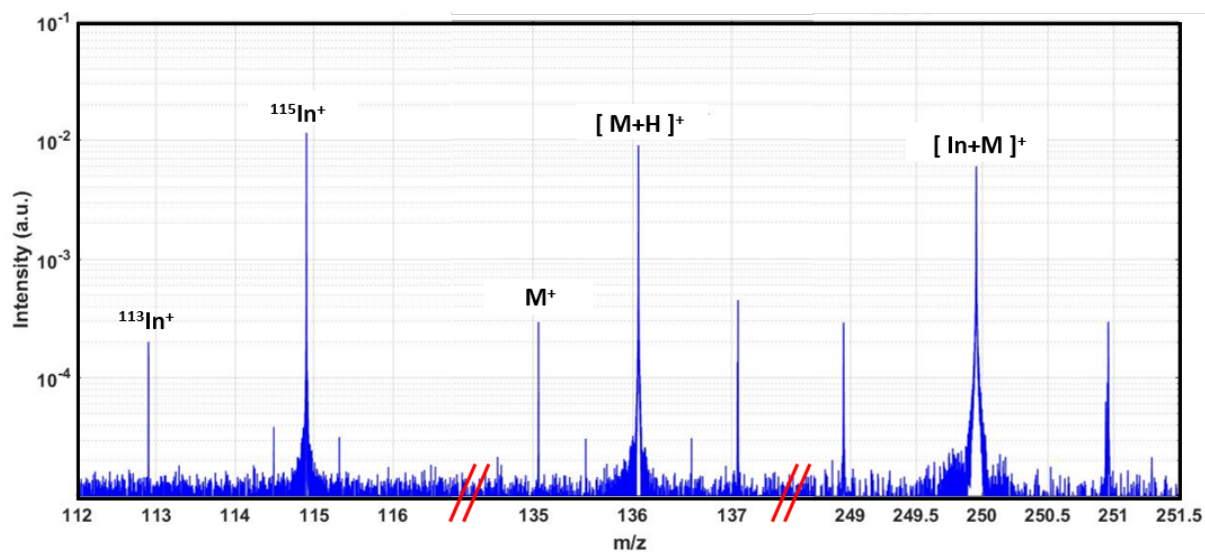

*Figure 3: Detail of a mass spectrum of adenine (M) powder pressed on an indium surface showing the formation of clusters between the sample-holder (indium) and adenine. The left mass window shows the indium main isotope peak $^{115}In^+$ and its isotope at m/z 113. The mass window in the middle details the molecular ion peak of adenine ($M^+$) and its protonated ion $[M+H]^+$. One mass unit higher, at m/z 137, we observe an isotopologue of the adenine protonated ion. The right mass window shows the cluster formed between the indium ion $^{115}In^+$ and the neutral adenine M. This $[In+M]^+$ peak is surrounded by two other peaks, close to the nominal m/z 249 and 251. Mass gaps with the cluster at m/z 250 are close to the mass of one hydrogen, for each one. Peaks at m/z 114.5, 115.3, 135.5, and 136.6 are artifacts due to the FFT treatment*

The presence of the indium peak is also an indicator of the laser beam position, as it disappears from the spectrum when the laser beam irradiates the selected organic samples which absorb less energy from the 266 nm UV laser in this setup. At the edge of the sample and the sample-holder, both are visible on the spectrum (Figure 3).

### c) Validation of the ionisation method for the blind test

Tests are performed on the sample with a progressive increase of the laser power from 30 µJ up to 750 µJ, until a complex specific pattern is observed in the spectrum. A systematic study of the mass spectra as a function of the laser power was not possible, due to the too small amount of sample available. One of the main features of laser ablation ionisation (LAb) with organic molecules is the fragmentation resulting from the dissociative ionisation of the sample. This can induce a high variability and diversity of mechanisms generating the mass spectrometric signature for organic molecules, which is not observed for metal, essentially resulting in singly charged atomic ions and clusters. Clusters are positively charged entities from the addition of one metallic ion coming from the surface of the sample-holder to a neutral organic molecule thermo-desorbed from the sample, as described in the previous section.

To ensure that this ionisation method is relevant for this blind test study and appropriate in the context of a space configuration, we compare the mass spectrometric signature of adenine obtained on the one hand by the well-documented electron ionisation process at 70 eV (NIST database) and on the other hand by the LAb technique of our prototype (Figure 4). Similarities in the fragmentation patterns are observed between both mass spectra. The top panel, in red, shows a LAb-CosmOrbitrap mass spectrum. The lower panel, in black and with a reverse y-scale, shows the electron ionisation mass spectrum, recovered from the NIST database. Intense peaks and parent molecules are identified on the mass spectra. Intensities of ions detected are different but main fragment ions are the same for both ionisation processes. Two of the most intense fragment ions are at *m/z* 28 and 108. The fragment ion at *m/z* 108 is interpreted as a loss of an HCN molecule from the adenine molecular ion observed at *m/z* 135.0527. Based on exact masses calculations, the *m/z* 108 fragment ion is expected at *m/z* 108.0431. We observed it at *m/z* 108.0427 (-3.7 ppm). Laser ablation and electron ionisation (EI) are no soft ionisation processes and produce molecular ions and subsequent fragmentations. The fragmentation results from molecular ions presenting an unstable structure due to (1) the loss of an electron from the electronic cloud and (2) the internal energy left in the ionic species, imparted either by the incident electron or the plasma generated by the laser. For a given molecular structure, the fragmentation routes that results from this process should be similar. This explains why the main fragment ions observed are similar with both ionisation methods. However differences of peak intensities are observed between the two mass spectra. This is the signature that the internal energy imparted in the ablation plasma is different from the one imparted by the 70 eV electrons. This is understandable as the laser generates micron size plasma, with a lifetime of hundreds of nanoseconds. This is a medium presenting, from the ion microphysical point of view, very diverse conditions. Hence, the ion can be generated in the middle of the plasma plume or close to its edges, or it can be generated at the beginning of the ablation process, in high density of charges and gaseous species, or later on, during the recombination steps following the photon extinction. This variability in time and position results in different amount of internal energy imparted to the primary ions, leading to different probability to further evolve into fragments. A further difference is in the repeatability of the spectra. Hence with electron ionisation, the energy used is constant, whereas the shot to shot laser variability is about 10%. Moreover, as we press solid powders onto indium, the surface of the sample is not homogenous and varies from one laser shot to another. This induces inherent variabilities among the LAb-CosmOrbitrap mass spectra.

The main difference in the LAb-CosmOrbitrap spectrum, compared to the NIST spectrum, is an important contribution at *m/z* 136. The MRP (FWHM) of 123,080 enables to identify the protonated ion of adenine $[M+H]^+$ at *m/z* 136.0614. It excludes the signature of $^{13}C$ and $^{15}N$ natural isotopologues of adenine at *m/z* 136.0574 and 136.0510, respectively, which would have been detected in the mass spectrum, in view of the mass resolution. The presence of this protonated ion results from proton addition by ion molecule reactions occurring inside the plume.

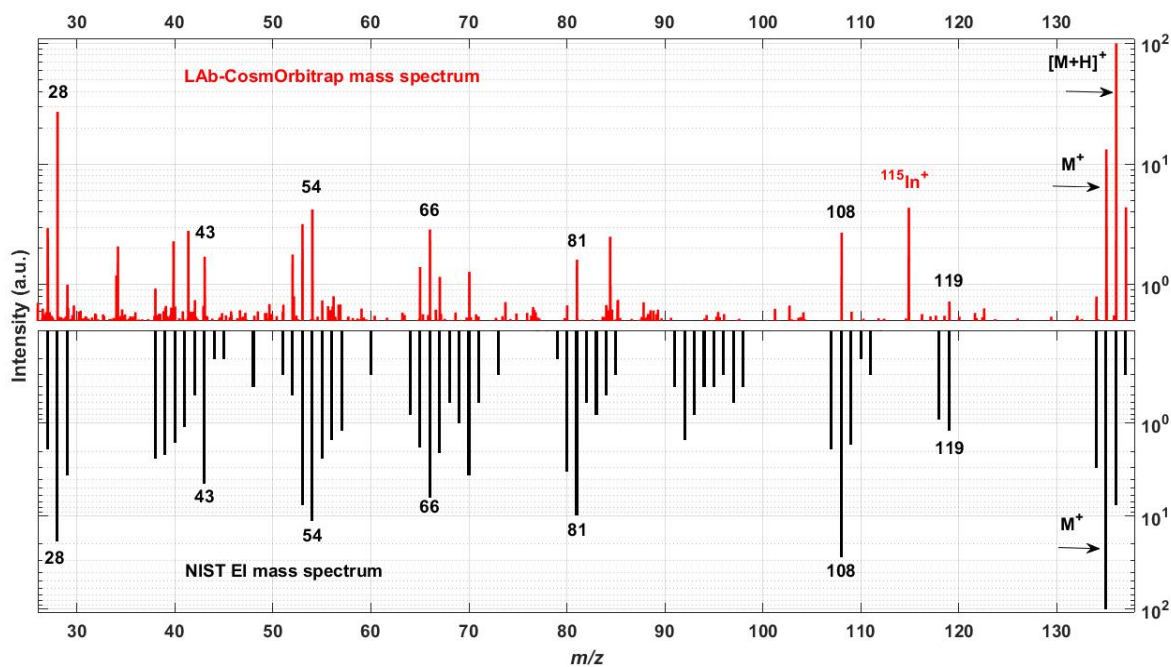

*Figure 4: Mass spectra of adenine obtained with LAb-CosmOrbitrap (red, upper panel) and from the NIST (black, lower panel, inverted scale) showing a comparison of the fragmentation patterns of adenine (M) between both ionisation processes: laser ablation and electron impact. On CosmOrbitrap spectrum, we see the molecular ion of adenine ($M^+$) and the protonated ion ($[M+H]^+$) both in positive mode. On the NIST spectrum, we observe only the molecular ion.*

The clear similarities observed between the two spectra of adenine allow us to conclude that adenine can be identified with the fragmentation patterns obtained in our experiment by comparison with its electron impact fingerprint provided in the NIST database. Note that obvious discrepancies are also observed, for instance in the *m/z* range 30 to 40 and 90 to 100. In our case, the applied laser power is only high enough to induce a fragmentation pattern. With a higher laser power, more fragment ions would be formed and detected. The electron impact ionization used to obtain mass spectra referenced in the NIST database involves a similar dissociating ionization process, breaking the molecular ion. For one given molecule, some fragments are produced with a higher probability which highly depends on the spatial configuration of the molecule and its chemical bonds. Electron impact ionization and laser ablation are preferentially producing the same fragments and it is why we observe the main fragment ions at the closest masses from the molecular ion in the mass spectra. However, the NIST mass spectrum shows more fragmentation peaks at lower masses than in our mass spectrum. This suggests that the electron ionization at 70 eV is more energetic than the photo-ionization process induced by the 266 nm laser used (corresponding to 4.6eV for one-photon transitions). The goal of this comparison (and of those which will be presented in the results section) is to confirm the attribution of a molecular formula and the identification of a compound by finding similar fragment ions between both techniques. We are not looking for two identical mass spectra but only a few and consistent similarities between both in order to confirm our identification. The fragmentation becomes a key asset and the final step in the identification process of a molecule and consequently the choice of the laser ablation as ionisation system has been found relevant for the blind test study.

### d) Data processing procedures for the blind test

Data processing of the spectra is made with the in-house Attributor software of HRMS analysis (described at https://frodsite.wordpress.com/reasearch/attributor/) developed in the Igor Pro environment. A Hann apodisation window and 3 zero padding (signal of 4 million points associated to 12 million points set to 0) are applied before a [6-838] ms FFT. Mass calibration is performed on the $^{115}$In$^+$ peak.

The study is based on two data representations: (1) the formal mass spectrum and (2) the "Mass Defect versus Exact Mass" diagram (MDvEM diagram). The latter represents the mass defect of each ion as a function of its exact mass. The mass defect (MD) of an element is the mass difference between its exact mass and its closest integer mass (Danger et al., 2013). By convention, the mass defect of Carbon ($MD_{Carbon}$) is set to 0. Indeed, its exact mass is 12 u, which already is the integer mass. If we consider the nitrogen element: its exact mass is 14.0031 u and the closest integer mass is 14 u. The difference between both gives a positive mass defect of +0.0031 u. On the contrary, indium shows a negative mass defect: the exact mass is 114.9038 u and the closest integer mass 115 u, yielding a negative mass defect of -0.0962 u. A valuable asset of this kind of diagram is the observation of specific trend lines. They are representative of repetitive molecular groups thus they act as molecular signatures. They allow us to make assumptions on the possible molecular groups and/or elements composing the molecule (as described in the results section and shown in Figure 6 & Figure 9). Artifact peaks (mostly some vertical alignments of points due to ringing phenomenon) are easily detected with this analytical representation, which allows to discard them.

Our identification steps, based on the study of the two data representations enounced and described above, are: (1) to find the parent peak, this means the *m/z* of the peak to identify; (2) to determine the molecular formula: for this, we make some assumptions on elements possibly composing the molecule and their occurrence (hydrogen and carbon are directly supposed to be present as we are looking for organic compounds but nitrogen, sulphur, oxygen or phosphorous can also be part of the molecular formula); (3) to confirm our identification based on the comparison of mass spectra from the NIST database (electron ionisation at 70 eV) and LAb-CosmOrbitrap mass spectra. This comparison is performed, for the selected species, by looking for a match of the same main fragment ions in both mass spectra (as described in the previous section).

The determination of the molecular formula (step 2) is helped by the Attributor software which calculates combinations of molecular formula at one *m/z* given. Each combination calculated is given with its mass accuracy (based on the difference between the theoretical *m/z* and the observed one) in ppm. Within this list of candidates, the most interesting ones (depending on their mass accuracy) are chosen and then the step 3 starts (comparison of the fragmentation pattern using the NIST database mass spectra as reference).

III) Results

a) Identification of "sample A"

The surface of "sample A" is scanned with successive laser shots at different locations on the sample-holder. A direct visualisation of mass spectra during the data acquisition allows to locate peaks in the *m/z* range 50 to 250, with some isolated peaks above *m/z* 350. Within this quite extended pattern, the identification of a parent molecule is not obvious. Figure 5 presents a representative mass spectrum. Among the most intense peaks detected, we observe the main indium isotope coming from the metallic surface of the sample holder at nominal *m/z* 115. Among the other intense peaks detected, nominal *m/z* 136, 250 and 364 are expected to be organic compound(s) or cluster(s) between indium and organics. In addition, in view of the large number of peaks detected, we assume a high fragmentation. This fragmentation is consistent with our laser ablation ionisation process which, as explained in the method section, induces it. Fragmentation comes from the molecular ion, then fragments ions produced can be themselves fragmented.

Between some clusters of peaks we are thus able to identify losses of only one element such as carbon, nitrogen, oxygen etc. These losses corresponding to only one element are good indicators for identifying elements composing the molecular formula. It is what we call "mass gaps": the mass difference between two peaks. The high MRP of the CosmOrbitrap (up to 90,000 at *m/z* 208 (lead) as referenced in the Briois et al., 2016 ) allows to calculate them with a precision of four digits in the *m/z* range 50 to 250.  As a reminder from the 2016 study, the CosmOrbitrap MRP (as for the conventional laboratory Orbitrap) is decreasing as a function of the square root of *m/z* (Briois et al., 2016; Makarov, 2000; Perry et al., 2008). The power law fitting this evolution of the MRP with the *m/z* ratio is given by: $m/\Delta m = k*(m/z)^{-1/2}$. Several nominal mass gaps of 12 u are detected (*m/z* 11.9999 for instance, attributed to carbon), but also 14 u (*m/z* 14.0028, attributed to nitrogen) and 16 u (*m/z* 15.9946, attributed to oxygen). Mass gaps of 31.9831 u are observed, which best correspond to losses (or addition) of two oxygen (mass = 31.9898 u for $O_2$). The high precision of this value enables to discard the presence of sulphur in "sample A" (*m/z* 31.7921). At this time, we are thus looking for a "CHNO" molecule.

The kind of mass spectrum presented on Figure 5 is usually observed at the edge of the sample deposit or after a series of shots, when the metal target starts to emerge. As mentioned in the method section, the ionisation of a solid sample deposited on a metal target can produce cluster ions such as [In+M]$^+$. Consequently, removing the mass of $^{115}$In to these cluster ions would lead to the neutral parent mass of sample "A" (molecule M). We obtain a nominal mass of 245 u when subtracting the mass of one $^{115}$In from the *m/z* 364 and a nominal mass of 135 u when subtracting the mass of one $^{115}$In from the *m/z* 250.

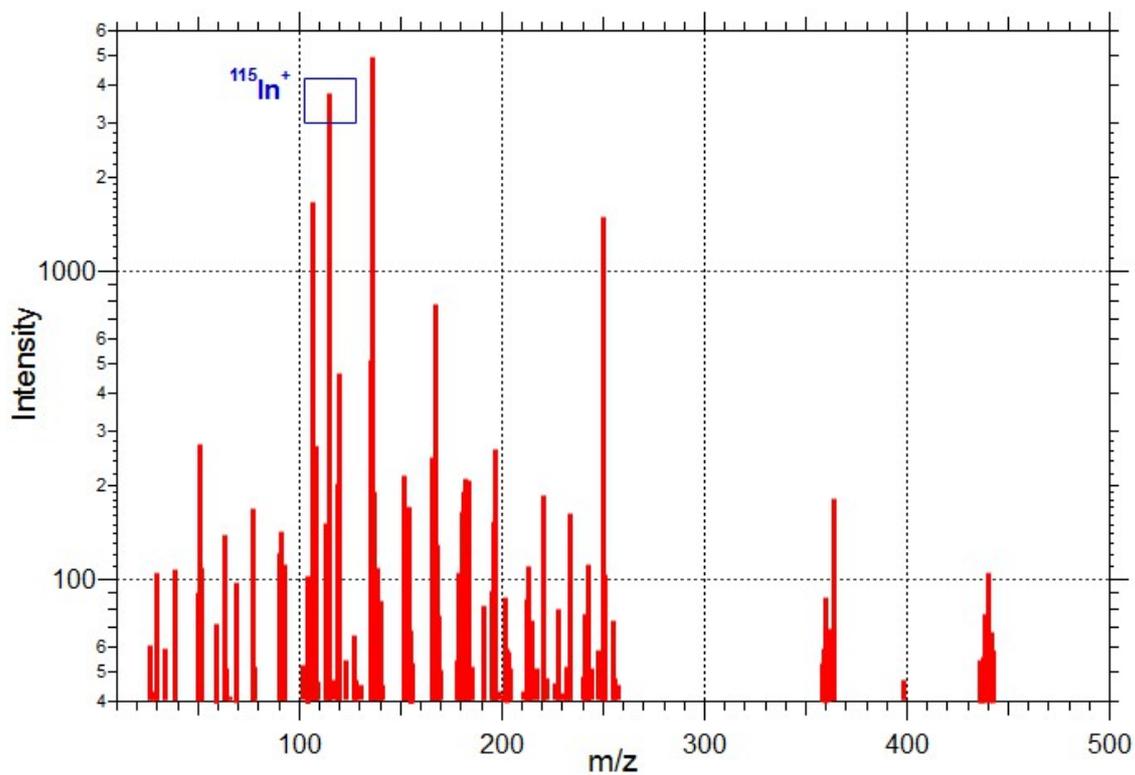

*Figure 5: Overall mass spectrum of sample "A". The blue square indicates the indium peak. Dynamic range is about 25 (based on $^{115}$In$^+$ and $^{113}$In$^+$).*

Further investigations using MDvEM diagram are performed with Attributor on the formal "sample A" mass spectrum as observed in Figure 5, in order to look for cluster ions. The MDvEM diagram is presented in Figure 6.

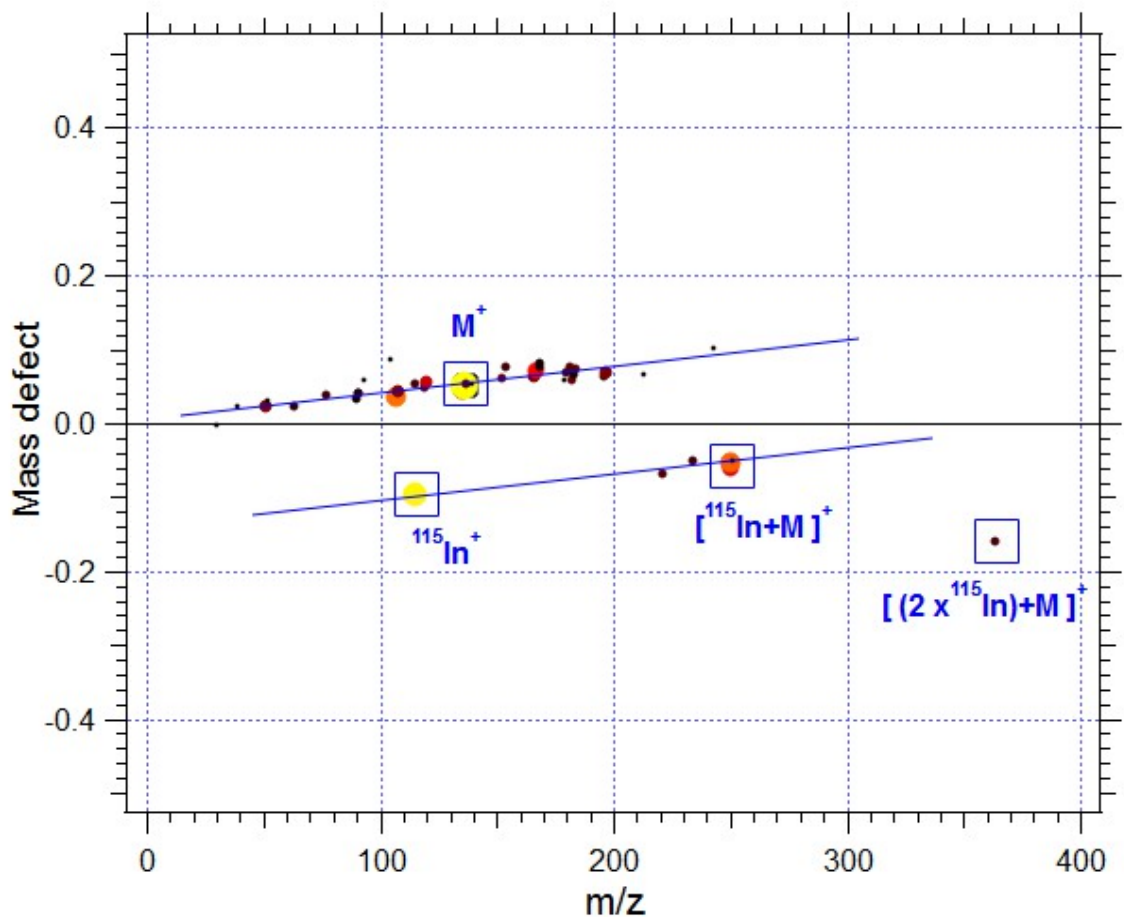

*Figure 6: Mass Defect versus Exact Mass (MDvEM) diagram of the « Sample A ». Each dot on the diagram corresponds to an ion detected in the mass spectrum (Figure 5). The mass defect is represented as a function of the m/z of each ion. In addition, colour and size of dots are proportional to the logarithm of the intensity of each ion peak. Blue lines highlight specific trend lines and cluster ions.*

Two different trends are evidenced. The first one (top blue line on Figure 6), with a positive MD, is attributed to organic species. Indeed, hydrogen and nitrogen elements have a positive MD (respectively +0.0078 and +0.0031), as explained in the method section carbon has a MD equal to 0 by convention and within these organic elements only the oxygen has a negative MD (-0.0051). Previously, we present mass gaps calculated between peaks on the *m/z* range 20 to 250 pointing toward the presence of a "CHNO" molecule with mainly C, H and N elements. We therefore expect a parent compound with a positive mass defect.

The second trend (bottom blue line with negative MD) is linked to the presence of indium, which has a MD of -0.0962. The yellow point corresponds to the large MD of indium alone, then the other points on this line corresponds to heteroatoms at higher masses: cluster ions of indium and the organic compound are suspected at $m/z$ 249.9455 ([In+M]$^+$) and at the nominal $m/z$ 365 [2In+M]$^+$.

From these cluster ions and as calculated previously in this section, we derive a hypothetical $m/z$ of the parent peak at 135.0417, consistent with the observed mass of a peak, located at $m/z$ 135.0426. With this laser ablation ionisation process and as demonstrated on adenine in the previous section we are observing both molecular and protonated ions. Thus, we assume that the peak at $m/z$ 135.0426 corresponds to the molecular ion and the peak at $m/z$ 136.0503 (more intense) to the protonated ion, with a mass gap between both of 1.0082 u, matching the mass of one hydrogen atom (1.0078 u). Then, we focus on the peaks observed at the nominal $m/z$ 137, to derive more information about the elements present in the molecule. We thus observe two adjacent peaks at this same nominal $m/z$: 137.0469 and 137.0537. The mass gap between both is 0.0068 u. Calculations based on their distance (in mass) from the protonated ion allow us to consider them both as isotopologues with the replacement in the molecular formula of the protonated ion of, respectively, one $^{14}$N by a $^{15}$N atom (first peak, with a mass gap of 0.9966 u) and one $^{12}$C by a $^{13}$C atom (second peak, with a mass gap of 1.0034 u). According to the nitrogen rule, the even $m/z$ of the protonated ion implies an odd number of nitrogen atoms. At the nominal $m/z$ 138 we do not observe a clear signal, confirming the absence of sulphur and a low abundance of oxygen. Based on all these assumptions (a CHNO molecule, a protonated ion at a $m/z$ close to 136.0503 u, the presence of an odd number of nitrogen atoms and oxygen in low abundance), 3 candidates are found compatible with a precision lower than 15 ppm on the exact mass attribution : $C_6H_6N_3O^+$ (-0.9960 ppm), $C_8H_8O_2^+$ (8.8018 ppm) and $C_4H_4N_6^+$ (-10.7940 ppm). The first possibility shows the lowest error. In addition, the presence of an oxygen loss with mass gap calculations excludes the third attribution. To confirm our choice, we look at the $C_6H_5N_3O$ molecule fragmentation pattern in the NIST database. We find that it is consistent with our spectrum (see Figure 7), mainly by the similar fragment ions at *m/z* 107 and 119. As for the comparison NIST/CosmOrbitrap made for adenine, all fragment ions are not comparable. More of them are detected on the NIST mass spectrum, obtained with a more energetic ionisation process. In this case, the *m/z* 119 fragment ion is relevant enough to confirm our molecule selection.

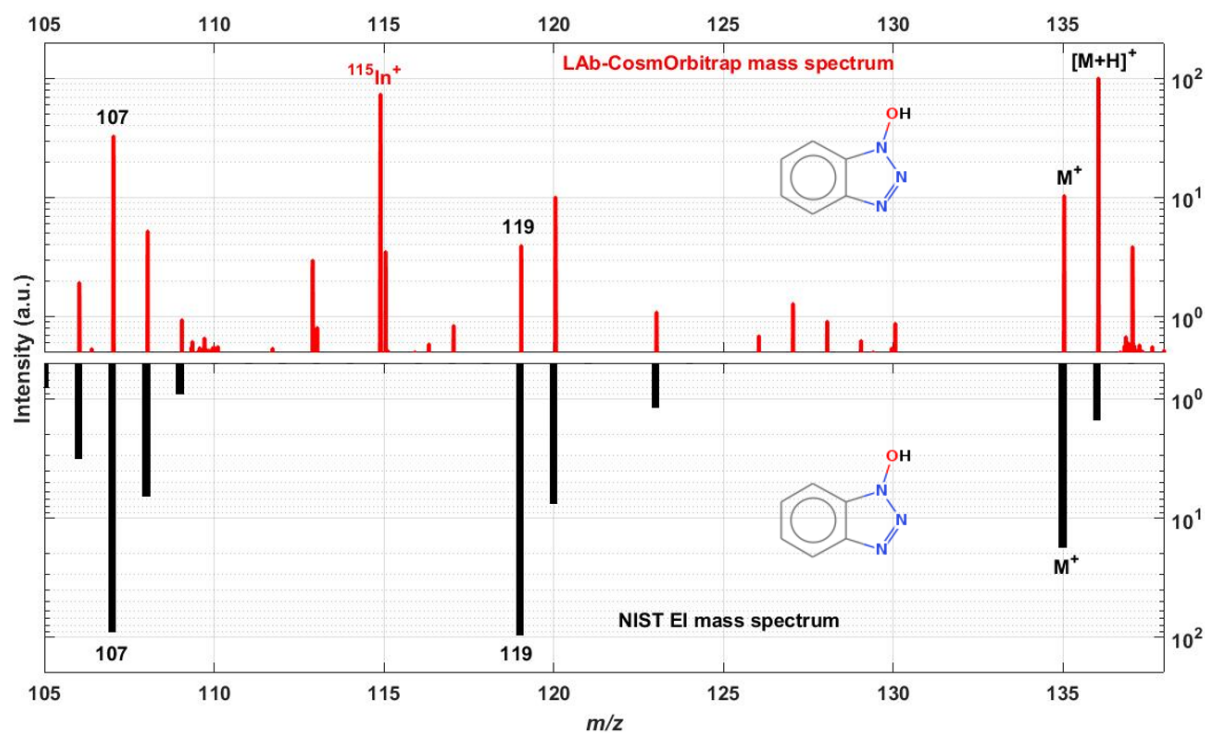

*Figure 7: Mass spectrum of sample "A" (assumed to be HOBt) obtained with LAb-CosmOrbitrap (top panel) and mass spectrum of HOBt from NIST database (lower panel). Comparison between both fragmentation patterns.*

Indeed, the identification steps detailed in this section lead to consider the sample "A" as HOBt, standing for 1H-Benzotriazole,1-hydroxy-, which is a derivative of benzotriazole. The presence of the benzotriazole fragment ion is thus very specific and detected in the CosmOrbitrap mass spectrum, at nominal *m/z* 119 (as well as in the NIST mass spectrum). The mass difference between the benzotriazole fragment ion and the molecular HOBt ion corresponds to the loss of an oxygen atom (15.9949 u). The MRP of the molecular ion at *m/z* 135 is 123,540. This organic compound is an unstable molecule when on its anhydrous form, which could explain the observed variability between mass spectra. The heating from the laser beam at the surface of the sample in addition to the ultra-high vacuum surrounding the sample yield an outgassing of water molecules from the sample, causing the studied HOBt compound to become unstable. Peaks between *m/z* 136 and 365 are interpreted as recombination between fragments of HOBt and HOBt or fragments of HOBt with indium (as observed for the cluster [indium + HOBt]$^+$ at *m/z* 250).

b) Identification of sample "B"

Using the same methodology, we obtain a mass spectrum for sample "B" with six specific and repeatable patterns from *m/z* 350 to 440 (Figure 8). The indium peak at the nominal *m/z* 115 is still visible (blue square in Figure 8, left spectrum), meaning the laser spot is at the edge of the sample deposit or the sample coverage at this point is thinner.

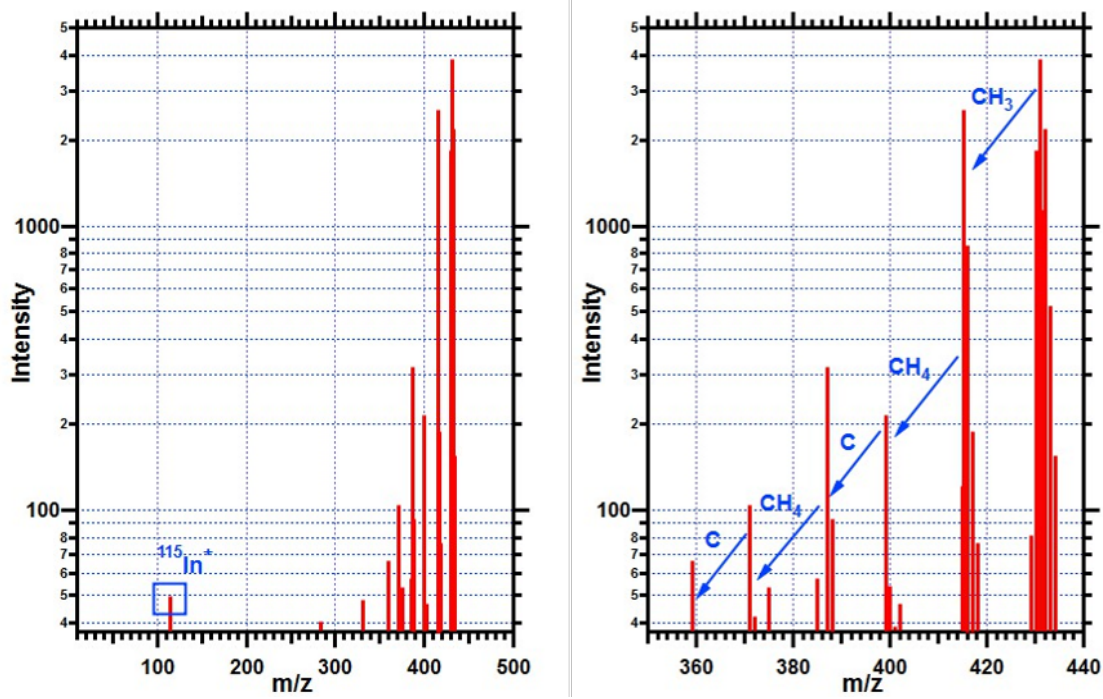

*Figure 8: (Left) Overall mass spectrum of sample "B" from 10 to 550 m/z range (left), where the blue square indicates the indium peak. (Right) Zoom on the six main patterns observed between 350 and 440 m/z range with the molecular formula of the lost fragments.*

Within this mass range, the most intense peak is located at *m/z* 431.1796. This observation is done during several consecutive laser shots, showing a strong repeatability. We consider this peak as the protonated ion of our molecule due to its mass difference of 1.0106 u, close to the mass of one hydrogen atom (1.0078), with the peak observed one nominal mass lower. Based on this assumption, the molecular ion should be the one located at *m/z* 430.1690. Its MRP is about 69,219.

Then, we focus on peaks detected from *m/z* 432 to 434. We interpret peaks at *m/z* 432.1813 and 433.1832 as isotopologues of the protonated ion, showing the replacement of respectively one $^{12}$C by one $^{13}$C and two $^{12}$C by two $^{13}$C in their molecular formula. The respective mass accuracy of these attributions are -0.7 and -4.2 ppm. The peak detected at *m/z* 434 should be an isotopologue with three $^{12}$C replaced by three $^{13}$C in its molecular formula. The decreasing intensities of these three peaks are consistent with the decreasing probabilities of carbon replacements in a given molecular formula. The *m/z* 434 isotopologue should be detected at *m/z* 434.1884. We indeed observe this value, but this is not the peak maximum, due to the bad shape of this peak. The high intensity of the nominal *m/z* 432 points to a large quantity of carbon atoms composing the molecular formula (around 30 carbon atoms to produce this signal intensity). At *m/z* 433 we observe a twin peak. One has already been interpreted as two carbon isotopes replacement. The other one has to be identified. The mass difference between this peak at *m/z* 433.1736 and the protonated ion is 1.9940 u. This could match with the mass difference between two isotopes of several elements: chlorine (1.9970 u between $^{35}$Cl and $^{37}$Cl), sulphur (1.9958 u between $^{32}$S and $^{34}$S), nickel (1.9954 u between $^{58}$Ni and $^{60}$Ni) and iron (1.9983 u between $^{56}$Fe and $^{58}$Fe). The theoretical abundances of $^{37}$Cl and $^{60}$Ni do not match with the intensity of the observed peak (almost 25% for $^{37}$Cl and 26% for $^{60}$Ni against a mean of 10% for the experimental peak, based on several spectra), discarding these elements. Concerning iron, a peak should have been observed also at *m/z* 432. Finally, $^{34}$S is the isotope presenting the best match in terms of mass difference and relative intensity from the protonated one (theoretical intensity of 4.5%). Based on the nitrogen rule, the odd mass of the protonated ion induces an even number of nitrogen elements in the molecule.

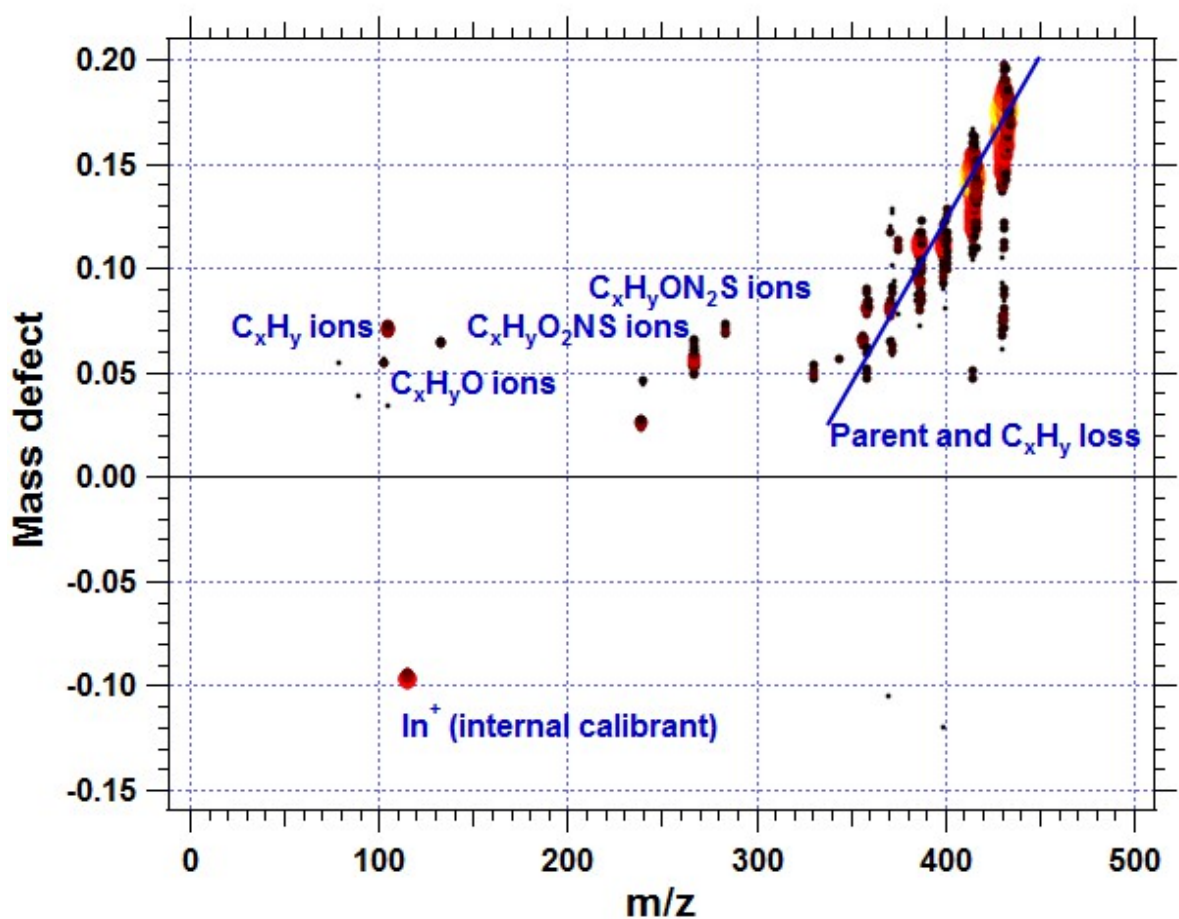

*Figure 9: Mass Defect versus Exact Mass (MDvEM) diagram of « Sample B ». Similarly to the MDvEM diagram of sample "A" (Figure 6), each ionic peak is represented as a dot corresponding to its mass defect as a function of its m/z, where, in addition, the colour and the size of the dots are proportional to the log of their intensities in the mass spectra. The blue line represents the main trend line observed (losses of $C_xH_y$ fragments). Other groups identified are indicated on the diagram. The dot corresponding to the indium ion is the largest one presenting a negative mass defect.*

The specific patterns at nominal *m/z* 415, 399, 387, 371 and 359 appear on the MDvEM diagram of Figure 9. These points are following the same trend, as a signature of the loss of the same molecular group, attributed in this case to specific losses of $C_xH_y$ from the molecular ion and proving the presence of abundant unsaturated chains in the structure of the molecule. These $C_xH_y$ fragments should be observed at lower masses. As we perform laser ablation, molecules are more and more broken at each laser shot at the same location. By this way, fragmentation increases with the number of laser shots. In order to derive more information from the MDvEM diagram, we study another spectrum obtained a few laser shots following the first one studied. On this mass spectrum, we observe the same six patterns from *m/z* 359 to 434 but also more fragment ions at lower intensities (as visible on Figure 9), coming from the fragmentation of species detected up to *m/z* 434. At intermediate masses this diagram allows us to identify also the presence of oxygen by the observation of $C_xH_yO$, $C_xH_yO_2NS$ and $C_xH_yON_2S$ ions signatures. These molecular formula are derived from assignments proposed by the Attributor software. Calculations based on the MDvEM diagram lead us to confirm some of them and particularly those given here. An example can be given based on the peaks observed at *m/z* 105.0706 and 133.0651. Both values, in term of *m/z* and mass defect have to be consistent together to confirm which molecular group is lost or added between these two peaks. In this case, the molecular group should be CO. This is indeed corresponding (1) in term of *m/z* (the experimental mass difference of 27.9949 u is very close to the theoretical one of 27.9949 u) and (2) in term of mass defect calculation where only the MD of oxygen is involved (as the MD of carbon is 0) with an experimental value of -0.0055 close the theoretical value of -0.0051.

Four candidates are found compatible within the precision of the spectrum: $C_{26}H_{27}N_2O_2S^+$ (-2.2 ppm), $C_{23}H_{31}N_2O_2S_2^+$ (5.6 ppm), $C_{20}H_{35}N_2O_2S_3^+$ (13.3 ppm), $C_{20}H_{27}N_6O_3S^+$ (14.5 ppm).

The first candidate, presenting the best mass accuracy, is also the one that explains the high intensity observed at *m/z* 432 (matching with a high number of carbon atoms). We decide to focus on this first molecule and to look at its fragmentation mass spectrum on the NIST database.

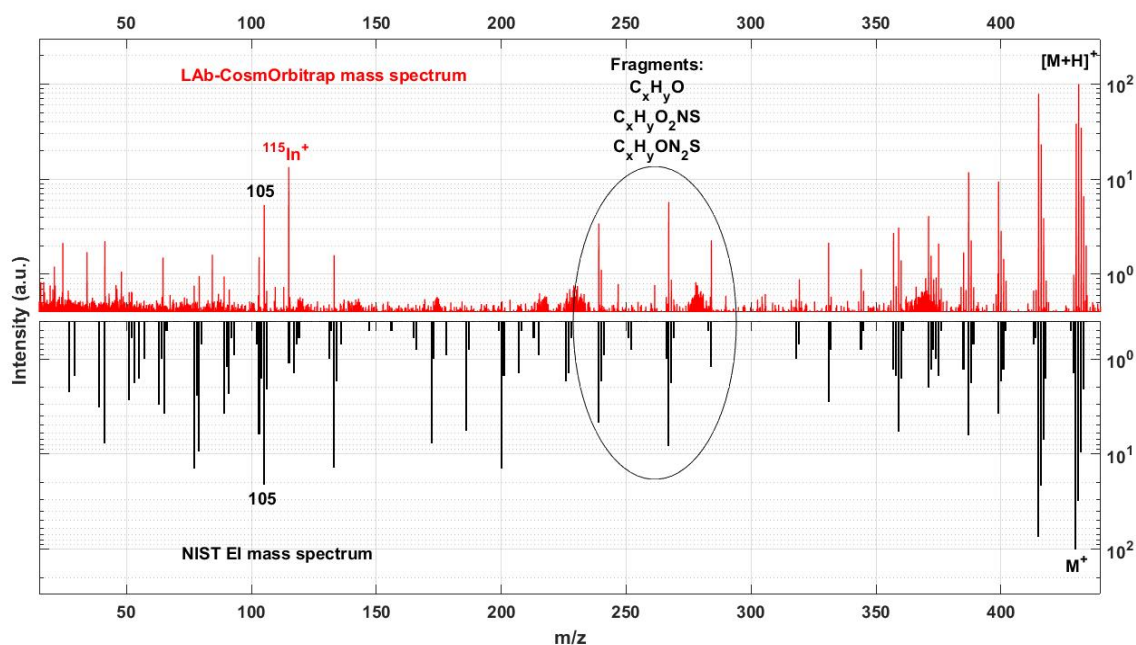

*Figure 10: Mass spectrum of sample "B" (assumed to be BBOT) obtained with LAb-CosmOrbitrap (top panel, in red) and mass spectrum of BBOT from NIST database (low panel in reverse scale, in black). Comparison of fragmentation patterns observed with these two techniques.*

As for the identification process of sample A, we compare both mass spectra on the same figure (Figure 10) with LAb-CosmOrbitrap mass spectrum on the top panel and NIST EI mass spectrum on the lower panel. The match between both spectra is mostly visible between *m/z* 350 and 431 but some intense fragments at the nominal *m/z* 103 and 267 are also observed. We select this molecule and identify it as BBOT, standing for 2,5-Bis(5-tert-butyl-2-benzoxazolyl)thiophene at a theoretical molecular and protonated m/z of, respectively, 430.1709 and 431.1787. The fragment ion at *m/z* 267.057 can be interpreted as $C_{15}H_{11}N_2OS$, which corresponds to a loss of $C_{11}H_{15}O$ consistent with the peak observed at this *m/z* on the MDvEM diagram (Figure 9). Some discrepancies are also visible on this mass spectra comparison but again, in this study we only focus on the main fragment ions produced by both techniques which allow us to identify the species.

## IV) Conclusion

Our blind analysis of the organic molecules A and B with the LAb-CosmOrbitrap prototype has led to the identification of respectively the two molecules HOBt (hydroxybenzotriazole, $C_6H_5N_3O$) and BBOT (2,5-Bis(5-tert-butyl-benzoxazol-2-yl)thiophene, $C_{26}H_{26}N_2O_2S$). Our attributions were afterwards positively confirmed by the JAXA HRMS team which had chosen and sent the samples.

Sample analysis and data treatment methodology require several steps. The process starts with a first visualisation of the full *m/z* range mass spectrum directly on the sample or at the edge of the sample and the metallic substrate, in order to add a specific feature by the production of cluster ions. This first step leads to the identification of the mass of the molecular ion. The high resolution of the spectrum enables attributions of the elements present in the molecule by calculation of mass gaps, within the precision of the spectra. To go deeper in the data treatment, the use of the mass defect versus exact mass (MDvEM) diagram allows to get information of possible repetitive patterns in the molecule to know more about the structure of the molecule and specific groups or atoms composing it. We can then infer the chemical formula of possible candidates. The comparison of the fragmentation patterns with the NIST EI (Electron Impact) spectra database finally confirms the selection of one molecule among the possible candidates.

This study shows the capability of the LAb-CosmOrbitrap instrument to identify unknown organic molecules. We also demonstrate that only a small amount of sample is needed to provide a good analysis. The instrument has an important potential for *in situ* chemical analysis of organic molecules in space, as molecular identification capability is absolutely required for new instrumental developments. Precious clues are given here to anticipate what is needed for a space configuration. We show the crucial importance of the sample preparation and the nature of the sample-holder. Deposition method on the surface of the sample-holder has to be set in order to benefit from the interaction between the substrate and the sample by clusters formation. Moreover, the metallic sample-holder gives us a useful calibration mass point for the whole mass spectrum. Finally, following the work by the Briois et al, 2016 and to update analytical performances of the LAb-CosmOrbitrap on organics, we show the capability of our instrumental configuration to reach a mass resolving power (MRP) of 69,219 at *m/z* 430 and 123,540 at *m/z* 135. Mass accuracies better than 3 ppm (-0.9 ppm for HOBt and -2.2 ppm for BBOT) have been demonstrated. The LAb-CosmOrbitrap is thus a key instrument for future space missions and particularly for organic worlds, with analytical performances never reached in space to this date.

However the study also points out some difficulties to be tackled in the future such as the variability resulting from the ionisation process chosen (laser ablation) but also from the intrinsic properties of the sample (UV absorption, ionisation potential). As a wide energy range can be applied on the sample, a specific Laser Ablation CosmOrbitrap calibration mass spectra database would be relevant in order to get reference mass spectra at different energies of organic species presenting a potential exobiological interest. This database would allow a post data treatment of the spectra received from a spacecraft. In addition, a laser system with variable output energy should be set for a space configuration, in agreement with present laboratory configurations.

Acknowledgements

L. S. thanks the CNES (Centre National d'Etudes Spatiales) and the Région Centre-Val de Loire for the funding of her PhD.

N.C. thanks the European Research Council for funding via the ERC PrimChem project (grant agreement No. 636829.)

We gratefully acknowledged CNES for funding the instrumental development of the CosmOrbitrap, the Labex Exploration Spatiale des Environnements Planétaires (ESEP), and the Région Centre-Val de Loire, Dr Titaina Gibert for lending us the Nd-Yag laser 266 nm used to perform all tests and the JAXA HRMS team for initiating this study.

Young, D.T., Berthelier, J.J., Blanc, M., Burch, J.L., Coates, A.J., Goldstein, R., Grande, M., Hill, T.W., Johnson, R.E., Kelha, V., Mccomas, D.J., Sittler, E.C., Svenes, K.R., Szegö, K., Tanskanen, P., Ahola, K., Anderson, D., Bakshi, S., Baragiola, R.A., Barraclough, B.L., Black, R.K., Bolton, S., Booker, T., Bowman, R., Casey, P., Crary, F.J., Delapp, D., Dirks, G., Eaker, N., Funsten, H., Furman, J.D., Gosling, J.T., Hannula, H., Holmlund, C., Huomo, H., Illiano, J.M., Jensen, P., Johnson, M.A., Linder, D.R., Luntama, T., Maurice, S., Mccabe, K.P., Mursula, K., Narheim, B.T., Nordholt, J.E., Preece, A., Rudzki, J., Ruitberg, A., Smith, K., Szalai, S., Thomsen, M.F., Viherkanto, K., Vilppola, J., Vollmer, T., Wahl, T.E., Wüest, M., Ylikorpi, T., Zinsmeyer, C., 2004. Cassini Plasma Spectrometer Investigation. Space Sci. Rev. 114, 1–112. https://doi.org/10.1007/s11214-004-1406-4

Website source:

Attributor software of HRMS analysis at https://frodsite.wordpress.com/reasearch/attributor/, last access April 2018